\documentclass[11pt,fleqn]{article}

\usepackage{amsmath,amssymb,amsthm,cite,enumerate}

\newcommand{\todo}[1][\null]{\ensuremath{\clubsuit}}

\newcommand{\noprint}[1]{}

\newcommand{\p}{\partial}
\newcommand{\const}{\mathop{\rm const}\nolimits}

\newcommand{\lsemioplus}{\mathbin{\mbox{$\lefteqn{\hspace{.77ex}\rule{.4pt}{1.2ex}}{\in}$}}}

\newcounter{tbn}

\newcounter{mcasenum}

\newtheorem{theorem}{Theorem}

\newtheorem{corollary}{Corollary}
\newtheorem{proposition}{Proposition}
\newtheorem*{proposition*}{Proposition}
{\theoremstyle{definition}

\newtheorem{remark}{Remark}
}

\begin{document}

\begin{center}{\LARGE\bf
Lie symmetries and exact solutions\\ of variable coefficient mKdV equations:\\[.7ex]
an equivalence based approach}

\vspace{4mm}
{\Large Olena Vaneeva}

\vspace{2mm}
{\it Institute of Mathematics of NAS of Ukraine, 3 Tereshchenkivska Str., Kyiv-4, 01601 Ukraine}

\vspace{2mm}
{\it vaneeva@imath.kiev.ua}
\end{center}

{\vspace{5mm}\par\noindent\hspace*{8mm}\parbox{146mm}{\small
Group classification of classes of mKdV-like equations with time-dependent coefficients is carried out.
The usage of equivalence transformations appears to be a crucial point for the exhaustive solution of the problem.
We prove that all the classes under consideration are normalized.
This allows us to formulate the classification results in three ways:
up to two kinds of equivalence (which are generated by
transformations from the corresponding equivalence groups and
all admissible point transformations) and using no equivalence.
A simple way for the construction of exact solutions  of mKdV-like equations
using equivalence transformations is described.
}\par\vspace{2mm}}

\section{Introduction}
In the last decade there is an explosion of research activity in the investigation of different generalizations of
the well-known
equations of mathematical physics such as the KdV and mKdV equations,
the Kadomtsev--Petviashvili equation, nonlinear Schr\"odinger equations, etc.
A number of the papers devoted to the study of variable coefficient KdV or mKdV equations with time-dependent coefficients
were commented in~\cite{Popovych&Vaneeva2010}. Common feature of the commented papers is that results were
obtained mainly for the equations which are reducible to the standard KdV or mKdV equations by point transformations.
Moreover we noticed that usually even for such reducible equations the authors do not use equivalence transformations and perform
 complicated calculations of systems involving a number of unknown functions using computer algebra packages.
It appears that the usage of equivalence transformations allows one to obtain further results in a simpler way.

The aim of this paper is to demonstrate this fact and to present the correct group classification of a class of variable coefficient
mKdV equations. Namely, we investigate Lie symmetry properties and exact solutions of variable coefficient
mKdV equations of the  form
\begin{equation}\label{vc_mKdV}
u_t+u^2u_x+g(t)u_{xxx}+h(t)u=0,
\end{equation}
where $g$ and $h$ are arbitrary smooth functions of the variable $t$, $g\neq0.$
It is shown in Section~2 that using equivalence transformations the function $h$ can be always set to the zero value and therefore
the form of $h$ does not affect results of group classification.
So, at first we carry out the exhaustive group classification of the subclass of class~\eqref{vc_mKdV} singled out by the condition $h=0$.
Then using the classification list obtained and equivalence transformations we present group classification of the initial
class~\eqref{vc_mKdV}.

Moreover, equivalence transformations appear to be powerful enough to present the group classification for much
wider class of variable coefficient mKdV equations
of the form
\begin{equation}\label{EqvcmKdV}
u_t+f(t)u^2u_x+g(t)u_{xxx}+h(t)u+(p(t)+q(t)x)u_x+k(t)uu_x+l(t)=0,
\end{equation}
where all parameters are smooth functions of the variable $t$, $fg\neq0$ and the parameters $f, h, k$ and $l$
satisfy the condition
\begin{equation}\label{condition_reduce}
2lf=k_t+kh-k\frac{f_t}f.
\end{equation}
This result can be easily obtained due to the fact that the group classification problem for class~\eqref{EqvcmKdV}
can be reduced to the similar problem for class~\eqref{vc_mKdV} with $h=0$ if and only if condition~\eqref{condition_reduce} holds.
Namely, equations~\eqref{EqvcmKdV} whose
coefficients satisfy~\eqref{condition_reduce} are transformed to equations from class~\eqref{vc_mKdV} with $h=0$
by the point transformations (see Remark 1 for details).
Equations
 from class~\eqref{EqvcmKdV} are important for applications and, in particular,
describe atmospheric blocking phenomenon~\cite{Tang&Zhao&Huang&Lou2009}.

An interesting property of the above classes of differential equations is that they are normalized,
i.e., all admissible point transformations within these classes are generated by transformations
from the corresponding equivalence groups.
Therefore, there are no additional equivalence transformations between cases of the classification lists,
which are constructed using the equivalence relations associated with the corresponding equivalence groups.
In other words, the same lists represent the group classification results for the corresponding classes
up to the general equivalence with respect to point transformations.

Recently the authors of~\cite{john10a} obtained a partial group classification of class~\eqref{vc_mKdV}
(the notation $a$ and $b$ was used there instead of $h$ and $g$, respectively.)
The reason of failure was neglecting an opportunity to use equivalence transformations.
This is why only some cases of Lie symmetry extensions were found, namely the cases with $h=\const$ and $h=1/t.$

In this paper we at first carry out the group classification problems
for class~\eqref{vc_mKdV} and
for the subclass of~\eqref{EqvcmKdV} singled out by condition~\eqref{condition_reduce} up to the respective
equivalence groups. (Throughout the paper  we use the notation$~\eqref{EqvcmKdV}|_\eqref{condition_reduce}$ for
the latter subclass.)
Then using the obtained classification lists and equivalence transformations we present group classifications of
classes~\eqref{vc_mKdV} and$~\eqref{EqvcmKdV}|_\eqref{condition_reduce}$
without the simplification of both equations admitting extensions of Lie symmetry algebras and these algebras themselves by equivalence transformations.
The extended classification lists can be useful for applications and convenient to be compared with the results of~\cite{john10a}.

Note that in~\cite{Magadeev1993,Gungor&Lahno&Zhdanov2004} group classifications for more general classes that include
 class~\eqref{EqvcmKdV} were carried out. Nevertheless those results obtained up to very wide equivalence group
  seem to be inconvenient to derive group classifications for classes~\eqref{vc_mKdV} and~\eqref{EqvcmKdV}.

In Section 5 we show how  equivalence transformations can be used to construct exact solutions for those
 equations from class~\eqref{EqvcmKdV} and its subclass~\eqref{vc_mKdV} which are
reducible to the standard mKdV equation.

\section{Equivalence transformations and mapping of class (1)\\ to a simpler one}

An important step under solving a group classification problem is the construction of the equivalence group
of the class of differential equations under consideration.
The usage of transformations from the related equivalence group often gives an opportunity to essentially simplify a
group classification problem and to present the final results in a closed and concise form.
Moreover, sometimes this appears to be a crucial point in
the exhaustive solution of such problems~\cite{IPS2007a,VJPS2007,VPS_2009}.

There exist several kinds of equivalence groups.
The \emph{usual equivalence group} of a class of differential equations consists of the nondegenerate point transformations
in the space of independent and dependent variables and arbitrary elements of the class
such that the transformation component for the variables do not depend on arbitrary elements
and each equation from the class is mapped by these transformations to equations from the same class.
If any point transformation between two fixed equations from the class belongs to its (usual) equivalence group then
this class is called \emph{normalized} in the usual sense.
See theoretical background on normalized classes in~\cite{Popovych2006c,Popovych&Kunzinger&Eshraghi2010}.

We find the equivalence
group $G^\sim_{1}$ of class~\eqref{vc_mKdV} using the results obtained in~\cite{Popovych&Vaneeva2010} for
more general class of variable coefficient mKdV equations.
Namely, in~\cite{Popovych&Vaneeva2010} a hierarchy of normalized subclasses
of the general third-order evolution equations was constructed.
The equivalence group for normalized class of variable coefficient mKdV
equations~\eqref{EqvcmKdV} (without restriction on values of arbitrary elements)
as well as criterion of reducibility of equations from this class to the standard mKdV equation were found therein.

The equivalence group $G^\sim$ of class~\eqref{EqvcmKdV} consists of the transformations
\begin{equation}\label{EqvcKdVEquivGroup}
\tilde t=\alpha(t),\quad
\tilde x=\beta(t)x+\gamma(t),\quad
\tilde u=\theta(t)u+\psi(t),\quad
\end{equation}
where $\alpha$, $\beta$, $\gamma$, $\theta$ and $\psi$ run through the set of smooth functions of~$t$, $\alpha_t\beta\theta\ne0$.
The arbitrary elements of~\eqref{EqvcmKdV} are transformed by the formulas
\begin{equation}\label{EqvcKdVEquivGroupArbitraryElementTrans}
\begin{split}&
\tilde f=\frac{\beta}{\alpha_t\theta^2}f, \quad
\tilde g=\frac{\beta^3}{\alpha_t}g, \quad
\tilde h=\frac1{\alpha_t}\left(h-\frac{\theta_t}\theta\right), \\&
\tilde q=\frac1{\alpha_t}\left(q+\frac{\beta_t}\beta\right), \quad
\tilde p=\frac1{\alpha_t}\left(\beta p-\gamma q+\beta\frac{\psi^2}{\theta^2} f-\beta\frac\psi\theta k+\gamma_t-\gamma\frac{\beta_t}\beta\right), \\&
\tilde k=\frac\beta{\alpha_t\theta}\left(k-2\frac\psi\theta f\right), \quad
\tilde l=\frac1{\alpha_t}\left(\theta l-\psi h-\psi_t+\psi\frac{\theta_t}\theta\right).
\end{split}
\end{equation}
The criterion of reducibility to the standard mKdV equation obtained in~\cite{Popovych&Vaneeva2010} reads as follows.

\begin{proposition}
An equation of form~\eqref{EqvcmKdV} is similar to the standard (constant coefficient) mKdV equation
if and only if its coefficients satisfy the conditions
\begin{equation}\label{EqvcmKdVEquivToKdV}
2h-2q=\frac{f_t}f-\frac{g_t}g, \quad
2lf=k_t+kh-k\frac{f_t}f.
\end{equation}
\end{proposition}

Class~\eqref{vc_mKdV} is a subclass of class~\eqref{EqvcmKdV} singled out by the conditions $f=1$ and $p=q=k=l=0.$
Substituting these values of the functions $f, p, q, k$ and $l$ to~\eqref{EqvcmKdVEquivToKdV} we obtain
the following assertion.

\begin{corollary}
An equation from class~\eqref{vc_mKdV} is reduced to the standard mKdV equation by a point transformation if and only if
\[
2h=-\frac{g_t}g,
\]
i.e. if and only if $g(t)=c_0\exp(-2\int h(t) dt),$ where $c_0$ is an arbitrary nonzero constant.
\end{corollary}

As class (2) is normalized~\cite{Popovych&Vaneeva2010}, its equivalence group $G^\sim$ generates
the entire set of admissible (form-preserving) transformations for this class.
Therefore, to describe of the set of admissible transformations for class~\eqref{vc_mKdV}
we should set $\tilde f=f=1,$ $\tilde p=p=\tilde q=q=\tilde k=k=\tilde l=l=0$ in~\eqref{EqvcKdVEquivGroupArbitraryElementTrans}
and solve the resulting equations with respect to transformation parameters.
It appears that projection of the obtained transformations on the space of the variables $t, x$ and $u$
can be applied to an arbitrary equation from class~\eqref{vc_mKdV}. It means that set of admissible transformations
of class~\eqref{vc_mKdV} is generated by transformations from its equivalence group and therefore this class is also
normalized.

Summing up the above consideration, we formulate the following theorem.

\begin{theorem}
Class~\eqref{vc_mKdV} is normalized.
The equivalence group~$G^{\sim}_1$ of this class consists of the transformations
\[
\begin{array}{l}
\tilde t=\beta\int\dfrac{dt}{\theta(t)^2},\quad
\tilde x=\beta x+\gamma, \quad \tilde u=\theta(t)u,\\[1ex]
\tilde h=\dfrac{\theta}{\beta}\left(\theta h-\theta_t\right), \quad
\tilde g=\beta^2\theta^2 g,
\end{array}
\]
where $\beta$ and $\gamma$ are arbitrary constants, $\beta\neq0$ and the function $\theta$ is an arbitrary nonvanishing smooth
function of the variable $t$.
\end{theorem}

The parameterization of transformations from the equivalence group $G^{\sim}_1$ by the arbitrary function $\theta(t)$
allows us to simplify the group classification problem for class~\eqref{vc_mKdV} via reducing the number of arbitrary elements.
For example, we can gauge arbitrary elements via setting either $h=0$ or $g=1$.
Thus, the gauge $h=0$ can be made by the equivalence transformation
\begin{equation}\label{gauge_h=0}
\tilde t=\int e^{-2\int h(t)\, dt}dt,\quad \tilde x=x, \quad
\tilde u=e^{\int h(t)\, dt}u,
\end{equation}
that connects equation~\eqref{vc_mKdV} with the equation
$\tilde u_{\tilde t}+\tilde u^2\tilde u_{\tilde x}+\tilde g(\tilde t){\tilde u}_{\tilde x\tilde x\tilde x}=0.$
The new arbitrary element $\tilde g$ is expressed via $g$ and $h$ in the following way:
\[
\tilde g(\tilde t)=e^{2\int h(t)\, dt}g(t).
\]

This is why without loss of generality we can restrict the study to the class
\begin{gather}\label{vc_mKdV_h=0}
u_t+u^2u_{x}+g(t)u_{xxx}=0,
\end{gather}
since all results on symmetries and exact solutions for this class can be extended to
class~\eqref{vc_mKdV} with transformations of the form~\eqref{gauge_h=0}.

The equivalence group for class~\eqref{vc_mKdV_h=0} can be obtained
from Theorem 1 by setting
$\tilde h=h=0$. Note that class~\eqref{vc_mKdV_h=0} is also normalized.

\begin{theorem} The equivalence group~$G^{\sim}_0$ of class~\eqref{vc_mKdV_h=0} is formed by the transformations
\[
\begin{array}{l}
\tilde t=\dfrac{\delta_2}{\delta_4^{\,2}} t+\delta_1,\quad \tilde x=\delta_2x+\delta_3,\quad
\tilde u=\delta_4u, \quad
\tilde g=\delta_2^{\,2}\delta_4^{\,2}g,
\end{array}
\]
where  $\delta_j,$ $j=1,\dots,4,$ are arbitrary constants,
$\delta_2\delta_4\not=0$.
\end{theorem}

\begin{corollary}
The equivalence algebra~$\mathfrak g^\sim$ of class~\eqref{vc_mKdV_h=0} is spanned by the operators
$\p_t$, $\p_x$, $t\p_t-\frac12u\p_u-g\p_g$ and $t\p_t+x\p_x+2g\p_g$.

\end{corollary}

\begin{remark}
An equation from class~\eqref{EqvcmKdV} is reducible to an equation from  class~\eqref{vc_mKdV_h=0} by a point transformation if and only if
its coefficients $f, h, k$ and $l$ satisfy the second condition of~\eqref{EqvcmKdVEquivToKdV},
i.e., condition~\eqref{condition_reduce}.
The corresponding transformation from~$G^{\sim}$ has the form
\begin{gather}\label{gauge2}
\begin{array}{l}\arraycolsep=0ex
\tilde t=\int fe^{-\int(q+2h)dt}dt,\quad \tilde x=e^{-\int q dt}x-
\int \left(p-\frac{k^2}{4f}\right)e^{-\int q dt}dt,\\
\tilde u=e^{\int h dt}\left(u+\tfrac k{2f}\right),\quad\tilde g=\dfrac gf e^{2\int(h-q)dt}.
\end{array}\end{gather}
In particular,  condition~\eqref{condition_reduce} implies that all equations from class~\eqref{EqvcmKdV}
with $k=l=0$ are reducible to equations from
class~\eqref{vc_mKdV_h=0}.
\end{remark}

\section{Lie symmetries}
We at first  carry out the group classification of class~\eqref{vc_mKdV_h=0} up to $G_0^\sim$-equivalence.
In this way we simultaneously solve the group classification problems
for class~\eqref{vc_mKdV} up to $G^\sim_1$-equivalence and
for the class$~\eqref{EqvcmKdV}|_\eqref{condition_reduce}$ up to $G^\sim$-equivalence
(see explanations below).
Then using the obtained classification lists and equivalence transformations we are able to present group classifications of
classes~\eqref{vc_mKdV} and$~\eqref{EqvcmKdV}|_\eqref{condition_reduce}$
without the simplification of equations with wider Lie invariance algebras by equivalence transformations.
These extended classification lists can be useful for applications and convenient to be compared with the results of~\cite{john10a}.

Group classification of class~\eqref{vc_mKdV_h=0} is carried out in the framework of the classical approach~\cite{Ovsiannikov1982}.
All required objects (the equivalence group, the kernel and all inequivalent cases of extension of the maximal Lie invariance algebras)
are found.

Namely, we look for the operators of the form $Q=\tau(t,x,u)\partial_t+\xi(t,x,u)\partial_x+\eta(t,x,u)\partial_u$,
which generate one-parameter groups of point symmetry transformations of equations from class~\eqref{vc_mKdV_h=0}.
These operators satisfy the necessary and sufficient condition of infinitesimal invariance, i.e.
action of the $r$-th prolongation~$Q^{(r)}$ of $Q$ to the ($r$-th order) differential equation (DE)
results in identical zero, modulo the DE under consideration. See,
e.g.,~\cite{Lahno&Spichak&Stognii2002,Olver1986,Ovsiannikov1982} for details. Here we require that
\begin{equation}\label{c1}
Q^{(3)}\big(u_t+u^2u_{x}+g(t)u_{xxx}\big)=0
\end{equation}
identically, modulo equation~\eqref{vc_mKdV_h=0}.

After elimination of $u_t$ due to~\eqref{vc_mKdV_h=0},
condition~(\ref{c1}) becomes an identity in eight variables, namely, the variables $t$, $x$, $u$, $u_x$, $u_{xx}$, $u_{tx}$, $u_{txx}$ and $u_{xxx}$.
In fact, equation~(\ref{c1}) is a
multivariable polynomial in the variables $u_x$, $u_{tx}$, $u_{xx}$, $u_{txx}$ and $u_{xxx}$. The
coefficients of the different powers of these variables must be
zero, giving the determining equations on the coefficients $\tau$, $\xi$ and $\eta$.
Since equation~\eqref{vc_mKdV_h=0} has a specific form
(it is a quasi-linear evolution equation,
the right hand side of~\eqref{vc_mKdV_h=0} is a
polynomial in the pure derivatives of $u$ with respect to $x$ etc),
the forms of the coefficients can be simplified.
That is, $\tau=\tau(t),$ $\xi=\xi(t,x)$~\cite{Kingston1991,Kingston&Sophocleous1998} and, moreover,
$\eta=\zeta(t,x)u$.
Then splitting with respect to~$u$ leads to the equations $\zeta_t=\zeta_x=0,$ $\xi_t=\xi_{xx}=0$,
$\tau_t-\xi_x+2\zeta=0$ and $\tau g_t=(3\xi_x-\tau_t)g$. Therefore,
we obtain the coefficients of the infinitesimal operator $Q$ in the form
\[
\tau=c_1t+c_2,\quad
\xi=c_3 x+c_4, \quad
\eta=\frac12(c_3-c_1)u,
\]
and the \emph{classifying} equation which includes arbitrary element $g$
\begin{equation}\label{Eqvc_mKdV_h=0ClassifyingEq}
\left(c_1t+c_2\right)g_t=(3c_3-c_1)g.
\end{equation}
The study of the classifying equation leads to the following theorem.
\begin{theorem}
The kernel~$\mathfrak g^\cap$ of the maximal Lie invariance algebras of equations from class~\eqref{vc_mKdV_h=0}
coincides with the one-dimensional algebra $\langle\partial_x\rangle$.
All possible $G_0^\sim$-inequivalent cases of extension of the maximal Lie invariance algebras are exhausted by the cases 1--3
of Table~1.
\end{theorem}
\begin{center}\small\renewcommand{\arraystretch}{1.6}
\setcounter{tbn}{-1}
\refstepcounter{table}\label{TableLieSymHF}
\textbf{Table~\thetable.}
The group classification of the class $u_t+u^2u_{x}+g\,u_{xxx}=0$, $g\neq0$.
\\[2ex]
\begin{tabular}{|c|c|l|}
\hline
N&$g(t)$&\hfil Basis of $A^{\max}$ \\
\hline
\refstepcounter{tbn}\label{TableLieSym_ker}\thetbn&$\forall$&$\partial_x$\\
\hline
\refstepcounter{tbn}\label{TableLieSym_2op}\thetbn&
$\delta t^n,\,n\neq0$&$\partial_x,\,6t\partial_t+2(n+1)x\partial_x+(n-2) u\partial_u$\\
\hline
\refstepcounter{tbn}\label{TableLieSym_3op}\thetbn&$\delta e^{t}$&
$\partial_x,\,6\partial_t+2x\partial_x+u\partial_u$\\
\hline
\refstepcounter{tbn}\label{TableLieSymHF_const}\thetbn&$\delta $&
$\partial_x,\,\partial_t,\,3t\partial_t+x\partial_x-u\partial_u$\\
\hline
\end{tabular}
\\[2ex]
\parbox{150mm}{Here $\delta=\pm1\bmod\, G^\sim_0,$ $n$ is an arbitrary nonzero constant.}
\end{center}

\begin{proof}
As class~\eqref{vc_mKdV_h=0} is normalized, it is also convenient to use a version of the algebraic method of group classification
or combine this method with the direct investigation of the classifying equation~\cite{DosSantosCardoso-Bihlo&Bihlo&Popovych2011}.
The procedure which we use is the following.
We consider the projection $\mathrm P\mathfrak g^\sim$ of the equivalence algebra~$\mathfrak g^\sim$
of class~\eqref{vc_mKdV_h=0} to the space of the variables $(t,x,u)$.
It is spanned by the operators $\p_t$, $\p_x$, $D^t=t\p_t-\frac12u\p_u$ and $D^x=x\p_x+\frac12u\p_u$.
For any~$g$ the maximal Lie invariance algebra of the corresponding equation from class~\eqref{vc_mKdV_h=0}
is a subalgebra of~$\mathrm P\mathfrak g^\sim$ in view of the normalization of this class
and contains the kernel algebra $\mathfrak g^\cap=\langle\partial_x\rangle$.
The algebra $\mathrm P\mathfrak g^\sim$ can be represented in the form
$\mathrm P\mathfrak g^\sim=\mathfrak g^\cap\lsemioplus\mathfrak g^{\rm ext}$,
where $\mathfrak g^\cap$ and $\mathfrak g^{\rm ext}=\langle D^t,D^x,\p_t\rangle$ is an ideal and a subalgebra
of~$\mathrm P\mathfrak g^\sim$, respectively.
Therefore, each extension of the kernel algebra $\mathfrak g^\cap$ is associated with a subalgebra of~$\mathfrak g^{\rm ext}$.
In other words, to classify Lie symmetry extensions in class~\eqref{vc_mKdV_h=0} up to $G_0^\sim$-equivalence it is sufficient
to classify $G_0^\sim$-inequivalent subalgebras of~$\mathfrak g^{\rm ext}$ and then check what subalgebras are agreed with
the classifying equation~\eqref{Eqvc_mKdV_h=0ClassifyingEq} and corresponds to a maximal extension.
The complete list of  $G_0^\sim$-inequivalent subalgebras of~$\mathfrak g^{\rm ext}$ is exhausted by the following subalgebras:
\begin{gather*}
\mathfrak g_0      =\{0\},\quad
\mathfrak g_{1.1}^a=\langle D^t+aD^x     \rangle,\quad
\mathfrak g_{1.2}^b=\langle D^x+b\p_t    \rangle,\quad
\mathfrak g_{1.3}  =\langle \p_t         \rangle,\quad
\mathfrak g_{2.1}  =\langle D^t,D^x      \rangle,\\
\mathfrak g_{2.2}^a=\langle D^t+aD^x,\p_t\rangle,\quad
\mathfrak g_{2.3}  =\langle D^x,\p_t     \rangle,\quad
\mathfrak g_3      =\langle D^t,D^x,\p_t \rangle,\quad
\end{gather*}
where the parameter $b$ can be scaled to any appropriate value if it is nonzero.
We fix a subalgebra from the above list and substitute the coefficients of each basis element of this subalgebra into
the classifying equation~\eqref{Eqvc_mKdV_h=0ClassifyingEq}.
As a result, we obtain a system of ordinary differential equations with respect to the arbitrary element~$g$.
The systems associated with the subalgebras $\mathfrak g_{1.2}^0$, $\mathfrak g_{2.2}^a$, where $a\ne1/3$, $\mathfrak g_{2.3}$
and $\mathfrak g_3$ are not consistent with the condition $g\ne0$.
The extensions given by the subalgebras $\mathfrak g_{1.3}$ and $\mathfrak g_{1.1}^{1/3}$
are not maximal since the maximal Lie invariance algebra in the case $g_t=0$ coincides with~$\mathfrak g_3$.
The subalgebras $\mathfrak g_0$, $\mathfrak g_{1.1}^a$, $\mathfrak g_{1.2}^b$ and $\mathfrak g_{2.2}^{1/3}$,
where $a\ne1/3$ and $b\ne0$,  correspond to cases~0, 1, 2 and~3, respectively. The parameter $n$ appearing
in case 2 is connected with the parameter $a$ via the formula $n=3a-1$, in case 3 the parameter $b$ is scaled to the value
$b=3.$
\end{proof}

For any equation from class~\eqref{vc_mKdV} there exists
an imaged equation in class~\eqref{vc_mKdV_h=0} with respect to transformation~\eqref{gauge_h=0}
(resp. in class$~\eqref{EqvcmKdV}|_\eqref{condition_reduce}$ with respect to
transformation~\eqref{gauge2}).
The equivalence group
$G_0^\sim$ of class~\eqref{vc_mKdV_h=0} is induced by the equivalence group~$G^\sim_1$ of class~\eqref{EqvcmKdV}
which, in turn, is induced by the equivalence group~$G^\sim$ of class~\eqref{EqvcmKdV}.
These guarantee that
Table 1 presents also the group classification list for class~\eqref{vc_mKdV} up to $G^\sim_1$-equivalence
(resp. for the class$~\eqref{EqvcmKdV}|_\eqref{condition_reduce}$ up to $G^\sim$-equivalence).
As all of the above classes are normalized, we can state that we obtain Lie symmetry classifications of these classes
up to general point equivalence.
This leads to the following corollary of Theorem~3.

\begin{corollary}
An equation from class~\eqref{vc_mKdV} (resp. class~\eqref{EqvcmKdV}) admits
a three-dimensional Lie invariance algebra if and only if
it is reduced by a point transformation to constant coefficient mKdV equation, i.e., if and only if $g(t)=c_0\exp(-2\int h(t) dt),$
where $c_0$ is an arbitrary nonzero constant (resp. if and only if conditions~\eqref{EqvcmKdVEquivToKdV} hold).
\end{corollary}

\begin{remark}[On contractions] There exists a connection between cases~1 and~2 of Table~1
which is realized via a limit process called contraction.
Examples of such connections
arising as limits between equations and their Lie invariance algebras are presented
in~\cite{Bluman&Reid&Kumei1988,Bluman&Kumei1989,Popovych&Ivanova2005PETs}. The precise definition and mathematical background for
contractions of equations, algebras of symmetries and solutions were first formulated in~\cite{IPS2007b}.
To make the limit process from case~1 to case~2, we apply equivalence transformation
$\mathcal T$: $\tilde t =n(t-1),$ $\tilde x=n^{1/3}x,$ $\tilde u=n^{-1/3}u$
to the  equation $u_t+u^2u_{x}+\delta t^n u_{xxx}=0$ (case~1 of Table~1), which results in the equation
$\tilde u_{\tilde t}+{\tilde u}^2\tilde u_{\tilde x}+\delta \left({\tilde t}/n+1\right)^n {\tilde u}_{\tilde x\tilde x\tilde x}=0.$
Then we proceed to the limit $n\rightarrow +\infty$ and obtain the equation
$\tilde u_{\tilde t}+{\tilde u}^2\tilde u_{\tilde x}+\delta e^{\tilde t} {\tilde u}_{\tilde x\tilde x\tilde x}=0$
(case~2 of Table~1). The same procedure allows one to obtain contraction between the corresponding Lie invariance algebras.
\end{remark}

To derive group classification of class~\eqref{vc_mKdV} which are not simplified by equivalence transformations,
we at first apply equivalence transformations from the group $G_0^\sim$ to the classification list presented in
Table 1 and obtain the following extended list:

\medskip

0. arbitrary $\tilde g\colon$  $\langle\partial_{\tilde x}\rangle$;

\smallskip

1. $\tilde g=c_0(\tilde t+c_1)^n\colon$  $\langle\partial_{\tilde x},\,6(\tilde t+c_1)\partial_{\tilde t}+2(n+1)
{\tilde x}\partial_{\tilde x}+(n-2){\tilde u}\partial_{\tilde u}\rangle$;

\smallskip

2. $\tilde g=c_0e^{m\tilde t}\colon$  $\langle\partial_{\tilde x},\,6\partial_{\tilde t}+2m{\tilde x}\partial_{\tilde x}
+m{\tilde u}\partial_{\tilde u}\rangle$;

\smallskip

3. $\tilde g=c_0\colon$  $\langle\partial_{\tilde x},\,\partial_{\tilde t},\,3\tilde t\partial_{\tilde t}+
{\tilde x}\partial_{\tilde x}-{\tilde u}\partial_{\tilde u}\rangle$.

\smallskip

Here $c_0$, $c_1$, $m$ and $n$ are arbitrary constants, $c_0m n\neq0$.

Then we find preimages of equations from class $\tilde u_{\tilde t}+\tilde u^2\tilde u_{\tilde x}+\tilde g(\tilde t){\tilde u}_{\tilde x\tilde x\tilde x}=0$
with arbitrary elements collected in the above list with respect to transformation~\eqref{gauge_h=0}.
The last step is to transform basis operators of the corresponding Lie symmetry algebras. The results are presented in Table~2.

\begin{center}\small
\setcounter{tbn}{-1}\renewcommand{\arraystretch}{1.6}
\refstepcounter{table}\label{TableLieSymHF2}
\textbf{Table~\thetable.}
The group classification of the class $u_t+u^2u_{x}+g\,u_{xxx}+h\,u=0$, $g\neq0$.
\\[2ex]
\begin{tabular}{|c|c|c|l|}
\hline
N&$h(t)$&$g(t)$&\hfil Basis of $A^{\max}$ \\
\hline
\refstepcounter{tbn}\label{TableLieSym_ker2}\thetbn&$\forall$&$\forall$&$\partial_x$\\
\hline
\refstepcounter{tbn}\label{TableLieSym_2op2}\thetbn&$\forall$&
$c_0 \left(\int e^{-2\int h\, dt}dt+c_1\right)^ne^{-2\int h dt}$&$\partial_x,\,H\partial_t+2(n+1)x\partial_x+(n-2-hH) u\partial_u$\\
\hline
\refstepcounter{tbn}\label{TableLieSym_3op2}\thetbn&$\forall$&$c_0 e^{\int\left( m e^{-2\int h\, dt}-2 h\right) dt}$&
$\partial_x,\,6e^{2\int h dt}\partial_t+2m x\partial_x+\left(m-6he^{2\int h dt}\right)u\partial_u$\\
\hline
\refstepcounter{tbn}\label{TableLieSymHF_const2}\thetbn&$\forall$&$c_0 e^{-2\int h dt}$&
$\partial_x,\,e^{2\int h dt}\left(\partial_t-hu\partial_u\right),\,H\partial_t+2x\partial_x-(2+hH) u\partial_u$\\
\hline
\end{tabular}
\\[2ex]
\parbox{150mm}{Here $c_0$, $c_1$, $m$ and $n$ are arbitrary constants, $c_0m n\neq0$,
and $H=6e^{2\int h dt}\big(\int e^{-2\int h dt} dt+c_1\big)$.
In case~3 $c_1=0$ in the formula for $H$.}
\end{center}
Now it is easy to see that Table 2 includes all cases presented in~\cite{john10a} as partial cases.

In a similar way, using transformations~\eqref{gauge2} we obtain group classification of
class$~\eqref{EqvcmKdV}|_\eqref{condition_reduce}$
without simplification by equivalence transformations.
The corresponding results are collected in Table~3.
\begin{center}\small
\setcounter{tbn}{-1}\renewcommand{\arraystretch}{2}
\refstepcounter{table}\label{TableLieSymHF2}
\textbf{Table~\thetable.}
The group classification of the class $u_t+f\, u^2u_x+g\,u_{xxx}+h\,u+(p +q\,x)u_x+k\,uu_x+\frac1{2f}\left(k_t+kh-k\frac{f_t}f\right)=0$,
$fg\neq0$.
\\[2ex]
\begin{tabular}{|c|c|l|}
\hline
N&$g(t)$&\hfil Basis of $A^{\max}$ \\
\hline
\refstepcounter{tbn}\label{TableLieSym_ker2}\thetbn&$\forall$&$e^{\int q dt}\partial_x$\\
\hline
\refstepcounter{tbn}\label{TableLieSym_2op2}\thetbn&
$c_0 fe^{2\int(q- h) dt}\left(\dfrac HF\right)^n$&$e^{\int q dt}\partial_x,\,H\partial_t+
\Bigl[(qH+2n+2)x+H\left(
p-\frac{k^2}{4f}\right)-$\\
&&$2(n+1)Q\Bigr]\partial_x+\left[(n-2-hH) u+\frac k{2f}(n-2)-lH\right]\partial_u$\\
\hline
\refstepcounter{tbn}\label{TableLieSym_3op2}\thetbn&$c_0 fe^{\int\left( m fe^{-\int(q+2h)\, dt}+2q-2 h\right) dt}$&
$e^{\int q dt}\partial_x,\,F\partial_t+\Bigl[(qF+2m)x+F\left(
p-\frac{k^2}{4f}\right)-2mQ\Bigr]\partial_x+$\\
&&$\left[(m-hF)u+\frac m2\frac kf-lF\right]\partial_u$\\
\hline
3&$c_0 fe^{2\int(q- h) dt}$&
$e^{\int q dt}\partial_x,\,F\Bigl[\partial_t+\left(qx+p-\frac{k^2}{4f}\right)\partial_x-
\left(hu+l\right)\partial_u\Bigr],\,H\partial_t+$\\
&&$
\Bigl[(qH+2)x+H\left(
p-\frac{k^2}{4f}\right)-2Q\Bigr]\partial_x-\Bigl[(2+hH) u+\frac k{f}+lH\Bigr]\partial_u$\\
\hline
\end{tabular}
\\[2ex]
\parbox{150mm}{The functions $f,$ $h$, $p$, $q$ and $k$ are arbitrary functions of the variable $t$ in all cases, $f\neq0.$

$c_0$, $c_1$, $m$ and $n$ are arbitrary constants, $c_0 m n\neq0$,
\begin{gather*}\begin{array}{l} F=\dfrac6f\,e^{\int(q+2 h) dt},\quad
H=F\left(\int fe^{-\int(q+ 2h)\, dt}dt+c_1\right),\\
Q=e^{\int{q dt}}\int\left(
p-\frac{k^2}{4f}\right)e^{-\int{q dt}} dt,\quad
l=\frac1{2f}\left(k_t+kh-k\frac{f_t}f\right).
\end{array}\end{gather*}
 In case 3 $c_1=0$ in the formula for $H$.}
\end{center}

\section{Construction of exact solutions using\\ equivalence transformations}
A number of recent papers concern the construction of exact solutions to different classes of KdV- or mKdV-like equations
using e.g. such methods as ``generalized $(G'/G)$-expansion method'', ``Exp-function method'',
``Jacobi elliptic function expansion method'', etc. A number of references are presented in~\cite{Popovych&Vaneeva2010}.
Moreover, we have noticed that usually authors of the above papers did not use equivalence transformations and carry out
complicated calculations for solution of systems involving a number of arbitrary functions using computer algebra packages.
Nevertheless, almost in all cases exact solutions were constructed only for equations which are reducible to the standard
KdV or mKdV equations by point transformations and usually these were only solutions similar to the well-known one-soliton solutions.
In this section we show that the usage of equivalence transformations allows one to obtain more results in a simpler way.

The $N$-soliton solution of the mKdV equation in the canonical form
\begin{equation}\label{canonical_mKdV}
U_t+6U^2U_{x}+U_{xxx}=0
\end{equation}
were constructed as early as in the seventies using the Hirota's method~\cite{Ablowitz&Segur}.
The one-  and two-soliton solutions of equation~\eqref{canonical_mKdV} have the form
\begin{gather}\label{sol1soliton}
U=a+\frac{k_0^2}{\sqrt{4a^2+k_0^2}\cosh z+2a}, \quad z=k_0x-k_0(6a^2+k_0^2)t+b,\\[2ex]\label{sol2soliton}
U=\frac{e^{\theta_1}\left(1+\dfrac{A}{4a_2^2}\,e^{2\theta_2}\right)+e^{\theta_2}\left(1+
\dfrac{A}{4a_1^2}\,e^{2\theta_1}\right)}{\left(
\dfrac1{2a_1}\,e^{\theta_1}+\dfrac1{2a_2}\,e^{\theta_2}\right)^2+\left(1-\dfrac{A}{4a_1a_2}\,e^{\theta_1+\theta_2}\right)^2},
\end{gather}
where $k_0, a, b, a_i, b_i$ are arbitrary constants, $\theta_i=a_ix-a_i^3t+b_i,$ $i=1,2;$ $A=\left(\dfrac{a_1-a_2}{a_1+a_2}\right)^2$.
Rational solutions which can be recovered by taking a long wave limit of soliton solutions
are also known for a long time~\cite{Ablowitz&Satsuma,Ono}.
Thus, the one- and two-soliton solutions give the rational solutions
\begin{gather}\label{sol_rational}
U=a-\frac{4a}{4a^2z^2+1}\quad\mbox{and}\quad
U=a-\frac{12a\left(z^4+\dfrac{3}{2a^2}z^2-\dfrac{3}{16a^4}-24tz\right)}{4a^2\left(
z^3+12t-\dfrac{3}{4a^2}\,z\right)^2+9\left(z^2+\dfrac1{4a^2}\right)^2},
\end{gather}
respectively, where $z=x-6a^2t$ and $a$ is an arbitrary constant.
These solutions can be found also in~\cite{Polyanin&Zaitsev}.
Note that solution~\eqref{sol2soliton} and the second solution of~\eqref{sol_rational} are presented in~\cite{Polyanin&Zaitsev} with misprints.

Combining the simple transformation $\tilde u=\sqrt6 U$ that connects the form~\eqref{canonical_mKdV}
of the mKdV equation with the form
\begin{equation}\label{mKdV_canonical}
\tilde u_{\tilde t}+{\tilde u}^2\tilde u_{\tilde x}+\tilde u_{\tilde x\tilde x\tilde x}=0
\end{equation}
and transformation~\eqref{gauge_h=0}, we obtain the formula
\[\textstyle  u=\sqrt{6}e^{-\int h(t)dt}\,U\left(\int e^{-2\int h(t)\, dt}dt,\,x\right).\]
Using this formula and solutions~\eqref{sol1soliton}--\eqref{sol_rational}
we can easily construct exact solutions for the equations of the general form
\begin{equation}\label{mKdV_canonical_preimage}
u_t+u^2u_{x}+e^{-2\int h\, dt}u_{xxx}+hu=0,
\end{equation}
which are preimages of~\eqref{mKdV_canonical} with respect to transformation~\eqref{gauge_h=0}.
Here $h=h(t)$ is an arbitrary nonvanishing smooth function of the variable~$t$.

For example, the two-soliton solution~\eqref{sol2soliton} leads to the following solution of~\eqref{mKdV_canonical_preimage}
\begin{gather*}
u=\sqrt{6}e^{-\int h\, dt}\frac{e^{\theta_1}\left(1+\dfrac{A}{4a_2^2}\,e^{2\theta_2}\right)+e^{\theta_2}\left(1+
\dfrac{A}{4a_1^2}\,e^{2\theta_1}\right)}{\left(
\dfrac1{2a_1}\,e^{\theta_1}+\dfrac1{2a_2}\,e^{\theta_2}\right)^2+\left(1-\dfrac{A}{4a_1a_2}\,e^{\theta_1+\theta_2}\right)^2},
\end{gather*}
where $a_i, b_i$ are arbitrary constants,
$\theta_i=a_ix-a_i^3\int e^{-2\int h\, dt}dt+b_i,$ $i=1,2$; $A=\left(\dfrac{a_1-a_2}{a_1+a_2}\right)^2$.
In a similar way one can easily construct one-soliton and rational solutions for equations from class~\eqref{mKdV_canonical_preimage}.

More complicated transformation of the form
\begin{gather*}\textstyle
u=\sqrt{6}e^{-\int hdt}\,U\left(\int fe^{-\int(q+2h)dt}dt,\,e^{-\int q dt}x-
\int \left(p-\frac{k^2}{4f}\right)e^{-\int q dt}dt\right)-\dfrac k{2f}
\end{gather*}
allows us to use solutions (11)--(13) of equation~\eqref{canonical_mKdV} for construction of exact
solutions of equations of the form
\begin{equation}\label{l_eq}
u_t+f u^2u_x+fe^{2\int(q-h)dt}u_{xxx}+h\, u+(p+q x)u_x+k\, uu_x+\frac1{2f}\left(\!k_t+kh-k\frac{f_t}f\!\right)=0,
\end{equation}
which  are preimages of equation~\eqref{mKdV_canonical} with respect to transformation~\eqref{gauge2}.
Here $f, h, k, p$ and $q$ are arbitrary smooth functions of the variable $t$, $f\neq0.$

For example, the solution of~\eqref{l_eq} obtained from the one-soliton solution (11) has the form
\begin{gather*}u=\sqrt{6}e^{-\int hdt}\left(a+\frac{k_0^2}{\sqrt{4a^2+k_0^2}\cosh z+2a}\right)-\dfrac k{2f}.
\end{gather*}Here
$z=k_0e^{-\int q dt}x-
k_0\int \left(p-\frac{k^2}{4f}\right)e^{-\int q dt}dt-k_0(6a^2+k_0^2)\int fe^{-\int(q+2h)dt}dt+b,$
where $k_0, a$ and $b$ are arbitrary constants.
 In a similar way one can easily construct two-soliton and rational solutions for equations from class~\eqref{l_eq}.

\section{Conclusion}
In this paper group classification problems for class~\eqref{vc_mKdV_h=0} and two more classes of variable coefficient
mKdV equations which are reducible to class~\eqref{vc_mKdV_h=0} by point transformations are carried out with respect
to the corresponding equivalence groups. Using the normalization property it is proved that these classifications coincide
with the ones carried out up to general point equivalence.
The classification lists extended by equivalence transformations are also presented. Such lists are convenient for applications.

It is shown that the usage of equivalence groups is a crucial point for exhaustive solution of the problem.
Moreover, equivalence transformations allow one to construct exact solutions of different types in a much easier way than by direct solving.
These transformations can also be utilized to obtain conservation laws, Lax pairs and other related objects
for equations reducible to well-known equations of mathematical physics by point transformations without direct calculations.

\subsection*{Acknowledgments}
The author thanks Prof. Roman Popovych for useful discussions and valuable comments.

\end{document}